\documentclass[prl,twocolumn,superscriptaddress]{revtex4-1}
\usepackage{amsmath}  % needed for \tfrac, \bmatrix, etc.
\usepackage{amsfonts} % needed for bold Greek, Fraktur, and blackboard bold
\usepackage{graphicx,color,calc,adjustbox}
\usepackage[version=4]{mhchem}% needed for figures
\usepackage{natbib} 
\usepackage[colorlinks=true,
urlcolor=black]{hyperref}
\definecolor{orange}{rgb}{1,0.5,0}

\usepackage[normalem]{ulem}
\usepackage {pifont}
\usepackage{color}
\begin{document}
\graphicspath{{./figs/}}
% \title{A physical solution to the crystallographic phase problem of diffraction for light}
%\title{Overcoming the phase problem of diffraction in photonic materials harnessing multi-wave interference}
%\title{Tackling the phase problem for determination of the spatial profile of nanocomposite photonic structures}
\title{Tackling the phase problem of diffraction for retrieval of photonic structures formed in nanocomposite materials}

\author{Martin Fally}
%\email{martin.fally@univie.ac.at}
\affiliation{Faculty of Physics, University of Vienna, 1090 Wien, Austria}

\author{Yasuo Tomita}
%\email{ytomita@uec.ac.jp}
\affiliation{Department of Engineering Science, University of Electro-Communications, 1-5-1 Chofugaoka, Chofu, Tokyo 182, Japan}

\author{Antonio Fimia}
%\email{a.fimia@umh.es}
\affiliation{Department of Material Science, Optics and Electronic Technology, University Miguel Hern\'{a}ndez, Elche, Alicante, Spain} 

\author{Roque F. Madrigal}
%\email{a.fimia@umh.es}
\affiliation{Department of Material Science, Optics and Electronic Technology, University Miguel Hern\'{a}ndez, Elche, Alicante, Spain}

\author{Jinxin Guo}%\email{jinxin.guo@bjut.edu.cn} 
\altaffiliation{Present address: Institute of Information Photonics Technology, Faculty of Science, Beijing University of Technology, Beijing 100124, China}
\affiliation{Department of Engineering Science, University of Electro-Communications, 1-5-1 Chofugaoka, Chofu, Tokyo 182, Japan}

\author{Joachim Kohlbrecher}
%\email{joachim.kohlbrecher@psi.ch}
\affiliation{Laboratory for Neutron Scattering, ETH Zurich \& Paul Scherrer Institut, 5232 Villigen PSI, Switzerland}

\author{J\"urgen Klepp}
\email{Corresponding author: juergen.klepp@univie.ac.at}
\affiliation{Faculty of Physics, University of Vienna, 1090 Wien, Austria}

% See the REVTeX documentation for more examples of author and affiliation lists.

\date{\today}

\begin{abstract}
We experimentally demonstrate how to solve the phase problem of diffraction using multi-wave interference with standard diffraction experimental setups without the need for taking any auxiliary data. In particular, we show that the phases of the Fourier components of a periodic structure can be fully recovered by deliberately choosing a probe wavelength of the diffracting radiation much smaller than the lattice constant. In the course of the demonstration, we accurately determine the refractive index profile of nanocomposite phase gratings by light and neutron diffraction measurements.
\end{abstract}

\maketitle 

Diffraction experiments are usually a method of choice to determine the internal structure of bulk materials.  
Instead of a crystal, let us consider here the simple case of a planar one-dimensional phase grating, which is characterized by the spatial profile of the refractive index
\begin{equation}\label{eq:RefrIndexPatt}
\tilde n(x)=\sum\limits_{s=-\infty}\limits^{+\infty} \tilde n_s e^{\imath s G x}.
\end{equation}
Here, $G$ is the spatial frequency, $|\tilde n_s|=n_s \in\mathbb{R}$ the amplitude and $\varphi_s=\arg (\tilde n_s)$ the relative phase of the Fourier component at the index $s$. In the case of phase gratings ($\tilde n_{-s}=\tilde n_{s}^* \in\mathbb{C}$) the Fourier-series of the real-valued refractive index reads $n(x)=n_0 + 2 \sum_{s=1}^{\infty} |\tilde n_s|\cos(sGx+\varphi_s)$. By determining the Fourier components, i.e. amplitudes and phases of the Fourier components, the structure is fully retrieved. However, by measuring the intensities of diffracted signals -- as it is usually done in standard diffraction experiments -- only the magnitude of the $s$-th Fourier component that corresponds to the $s$-th diffraction order, can be obtained, whereas the phases $\varphi_s$ are lost. The latter is called the phase problem of crystallography or diffraction (see, for instance, \cite{hauptmanRepProgPhys1991}). In principle, the phase problem can be overcome by measuring the phase differences between the wave incident to a sample of interest and each diffracted wave interferometrically, which, however, is found to be utterly difficult in many relevant cases. 

Sophisticated techniques have been developed to solve the phase problem and recover the phase information \cite{hauptmanRepProgPhys1991,woolfsonBook1995} either by so-called non-physical methods (direct methods, isomorphous replacement and anomalous scattering) or by introducing -- sometimes demanding -- experimental procedures to provide a physical solution \cite{chapmanPRL1981,changPRL1982,shenPRL1998,wolfPRL2009}. Discussions on available techniques and their development are ongoing (see, for instance, \cite{kleywegtActaCrystD2000,
cowtanELS2003,millaneActaCrystA2017,
fallySPIE2019Alt,donatelliPRL2020,heacockIUCrJ2020}). 
Surprisingly, sticking to traditional diffraction experiments alone can indeed provide all necessary information in many cases if experimental settings are chosen carefully, as we will show here.
In particular, we demonstrate by means of simple examples -- one-dimensional, non-sinusoidal phase gratings -- how such a physical solution to the phase problem of diffraction can be obtained with not more than the usual experimental effort in light and neutron diffraction experiments. The technique is applicable to samples generating diffraction patterns for any type of radiation useful for structure determination (X-rays, electrons, laser light, and neutrons, for instance). We apply our approach to the problem of determining deviations of the periodic refractive index profiles of holographic nanoparticle-polymer composite (NPC) gratings \cite{tomitaJOMO2016} from their ideal sinusoidal form. The information obtained is useful for our understanding of holographic formation processes \cite{tomitaSPIE2005,chikamaJAP2008} and development of versatile holographic applications such as diffractive optical elements, holographic data storage and display technology \cite{ohePAT1999,coufalHolDatSto2000,crawfordOPN2003,anIEEEPTL}.

Holographic gratings were prepared from a photopolymerizable NPC material. As reported in the past \cite{suzukiAO2004}, \ce{SiO2} nanoparticles (with the average diameter of 13 nm and the bulk refractive index $n_n$ of 1.46) dispersed in a solution of methyl isobutyl ketone are mixed with methacrylate monomers, 2-methyl-acrylic acid 2-4-[2-(2-methyl-acryloyloxy)-ethylsulfanylmethyl]-benzylsulfanyl-ethyl ester (the formed polymer refractive index $n_p$ is 1.59 at 589 nm). The doping concentration of \ce{SiO2} nanoparticles was 34 vol.\%. Photoinitiator titanocene (Irgacure 784, Ciba) is also mixed in 1 wt.\% with respect to the monomer to provide photosensitivity in the green. The above chemical mixture is cast on a glass plate. It is dried in an oven and finally covered with another glass plate, separated from the first one by spacers of known thickness. We employ a two-beam interference setup to write an unslanted transmission NPC grating by superposition of two mutually coherent s-polarized beams of equal light intensities from a laser diode-pumped frequency-doubled Nd:\ce{YVO4} laser operating at 532 nm. Two NPC gratings were prepared: G1 at grating spacing $\Lambda$ of $5\,\mu$m (with $\Lambda=2\pi/G$, see Eq.\,\ref{eq:RefrIndexPatt}) and with the thickness $d$ of about $13\,\mu$m, and G2 at spacing $\Lambda$ of $1\,\mu$m and with $d\approx 50\,\mu$m. In the bright regions of the interference pattern irradiating the sample, the photoinitiator triggers photopolymerization. Monomer is consumed there by the formation of polymer. Due to the resulting chemical potential gradient, a mutual diffusion process of unreacted monomer and nanoparticles sets in, leaving the dark regions enriched with nanoparticles \cite{tomitaJOMO2016}.  
It is important to note that, while the interference pattern is sinusoidal, the phase grating will be non-sinusoidal with nonzero higher Fourier components ($|s|>1$ in Eq.\,\ref{eq:RefrIndexPatt}) due to the interplay of nonlinear processes governed by the photopolymerization-driven mutual diffusion process for the formation of spatial density modulations of the formed polymer and nanoparticles. (Such characteristic features of the nanoparticle density profile can also be assessed by various theoretical models \cite{karpovOC2000,tomitaSPIE2005}.)
Thus, however elaborate the described production process might be, it is subject to continuing optimization and fine-tuning of experimental parameters (such as chemical compositions, nanoparticle concentration, and recording intensity and time) for the particular application one has in mind for the produced grating. Therefore, information as accurate as possible about the process outcome is desired, i.e. information about the redistribution of the nanoparticles dispersed in monomer under holographic exposure, which would eventually determine the exact form of the periodic refractive index profile given by Eq.\,\ref{eq:RefrIndexPatt}. It is the latter we investigate by our proposed method here.

\begin{figure}
\begin{center}
 \includegraphics[width=8.cm]{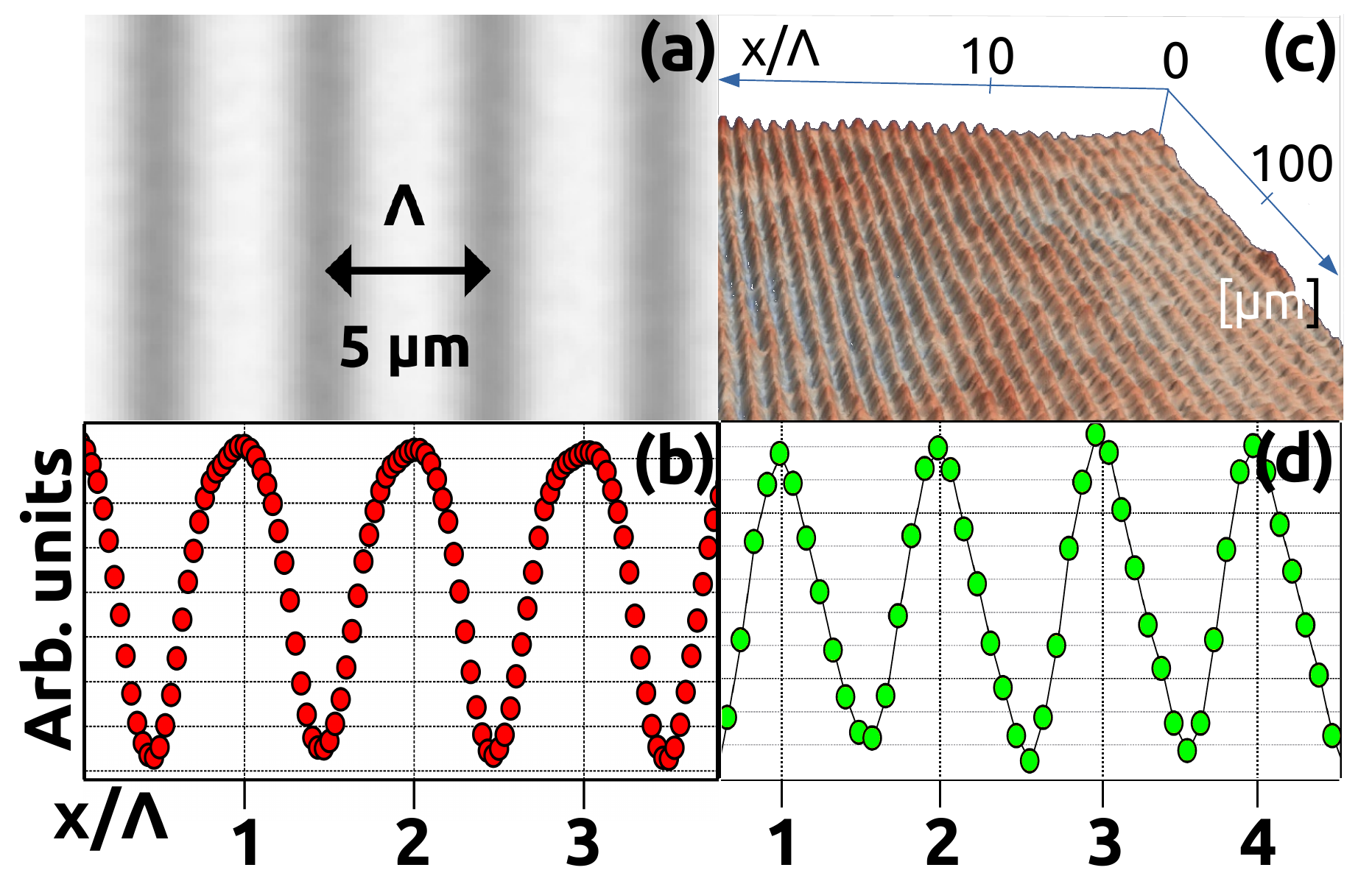}
\end{center}
\caption{\label{fig1} (a) Optical micrograph of the holographic \ce{SiO2} nanoparticle dispersed NPC grating G1 in transmission mode. (b) Spectral transmittance distribution along the grating vector of G1 obtained by averaging over lines of a rectangular region in Fig.\,\ref{fig1}\,(a) of several microns height. (c) Two-dimensional phase distribution of a grating similar to G1 measured by digital holographic microscopy. (d) Phase distribution of the grating along the grating vector obtained by averaging over 100 line scans in Fig.\,\ref{fig1}\,(c).  It is clearly seen that both profiles [(b) and (d)] exhibit higher order Fourier components.} 
\end{figure}
%First, we discuss G1, with longer $\Lambda$, for demonstrating our approach by comparison of the result to optical microscopy data.
Figure\,\ref{fig1}\,(a) shows an optical micrograph of G1 taken close to the maximum contrast position of the Talbot carpet 
\cite{goodmanBook2017,zhouAO2010,wenAOP2013}, which can be observed for pure phase gratings using an optical microscope. It can be seen that \ce{SiO2} nanoparticles (narrow, dark fringes, corresponding to lower refractive index regions) and the formed polymer (wide, bright fringes, corresponding to higher refractive index regions) are periodically arranged as a result of holographic assembly of nanoparticles \cite{tomitaOL2005}. Their distribution, shown in Fig.\,\ref{fig1}\,(b), is not perfectly sinusoidal. In Fig.\,\ref{fig1}\,(c), a section of a phase map of a sibling of G1 (also at $\Lambda=5~\mu$m) measured by digital holographic microscopy (DHM-R2100 by Lync\'{e}e Tec) is shown. The measurements were made with a 20X objective at a wavelength of $\lambda=684.9$\,nm. Also in this case, it is clearly seen from the sum of about 100 lines taken along the grooves of the phase map that the pattern is not purely sinusoidal, as shown in Fig.\,\ref{fig1}\,(d). Thus, the Fourier series describing both distribution patterns and, therefore, the refractive-index profiles must certainly include higher order Fourier terms, at $|s|>1$. 

\begin{figure}
\begin{center}
 \includegraphics[width=8.7cm]{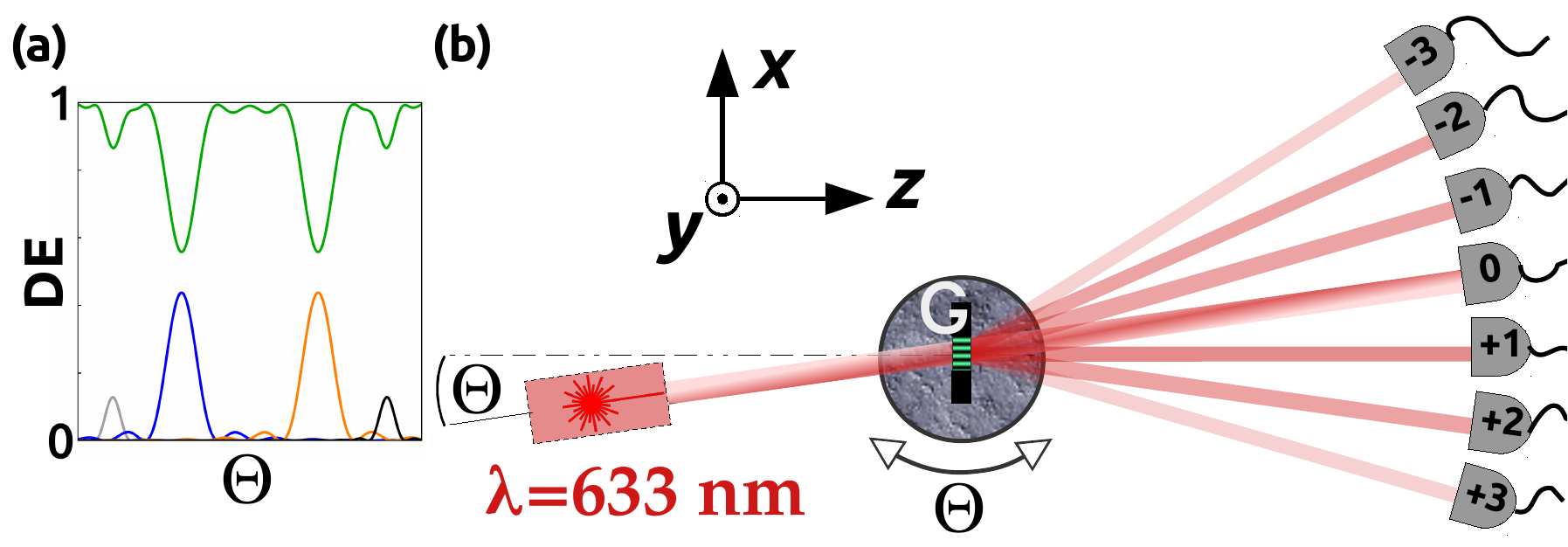}
\end{center}
\caption{\label{fig2} 
(a) Plot of typical DE in the Bragg regime as a function of angle of incidence $\Theta$. (b) Multi-wave interference data of the grating G1 are obtained diffracting a He-Ne laser of a few mm beam width at a wavelength of $\lambda=633$\,nm, much shorter than the grating spacing of $5\,\mu$m. A grating is placed on a motorized stage which is rotated about the $y$-axis.} 
\end{figure}

In many diffraction experiments aiming at structure determination, diffraction by a sample occurs in Bragg regime \cite{moharamOC1980}: At most two waves propagate and interfere simultaneously within the sample at all angles of incidence. 
One typically measures the dependence of the diffracted intensities upon the sample's rotation through angles of incidence $\Theta$ on a screen or a detector. In Bragg regime, peaks of the diffraction efficiency (DE, $\eta$) plotted versus $\Theta$ are relatively sharp and do not overlap as can be seen in Fig.\,\ref{fig2}\,(a). Interference of the two waves propagating within the sample's periodic structure (the refractive index modulation given by Eq.\,\ref{eq:RefrIndexPatt}) results in energy exchange between diffraction order pairs (the zero order beam and only one of the $\pm 1, \pm 2, \pm 3,\dots$ orders) as a function of $\Theta$. The first Born approximation does not hold in such a case, but theories such as dynamical diffraction theory or Kogelnik's theory (see, for instance, \cite{battermanRMP1964,kogelnikBSJ1969}) can be deployed for modelling data taken in Bragg regime. However, the relative phases $\varphi_s$ of Fourier components cannot be retrieved since multi-wave interference of the corresponding waves does not occur at any $\Theta$.
%\blue{To gain intuitive understanding as to why multi-wave interference is useful for phase retrieval, one may picture a structure diffracting in Bragg regime being analogous to a Mach-Zehnder interferometer, in which only two waves propagate and, depending on the geometry and interactions along the two paths, their interference results in the amplitude and phase of the wave leaving the interferometer. The latter phase (analogous to the phase of the outgoing waves in a diffraction experiment) cannot be measured unless we superpose the outgoing wave with a reference wave. A reference wave can, for instance, be obtained by replacing the first beamsplitter of the interferometer by a three-port beamsplitter and redirect the resulting third wave to the exit beamsplitter of the two-path device. Likewise, in a diffraction experiment it needs at least a third wave to be superposed with the wave resulting from the interference of the first two to obtain both amplitude and phase of a certain Fourier component.}
Now, in order to achieve such multi-wave interference (also referred to as multi-wave coupling) in diffraction and, thereby, to determine the relative phases $\varphi_s$ of G1's structure, we apply visible light of wavelength $\lambda\ll\Lambda$ as a probe in our diffraction experiment, such that $\Lambda$ is about an order of magnitude longer than the wavelength of the probe beam. 
By intentionally choosing $\lambda$ much shorter than it would seem appropriate, we can leave the Bragg regime behind and allow for multi-wave interference to occur within the periodic structure of the sample. 
From the Bragg equation $s\lambda=2\Lambda_s\sin\Theta_s$ ( $\Theta_s$ is the Bragg angle for the Fourier component at the index $s$) it is clear that the diffraction angle decreases with a decrease in $\lambda$, so that diffraction peaks of a given width overlap and the experiment can no longer be described in the Bragg regime. Clearly, more than just two waves propagate and interfere inside the sample at angles $\Theta$ for which observed peaks overlap considerably [see Fig.\,\ref{fig2}\,(b)].

The angular dependence data obtained by multi-wave interference diffraction can only be modelled by multiwave analysis theories such as the rigorous coupled-wave analysis (RCWA) \cite{moharamJOSAA1995}. Here, the strategy is to solve Maxwell's equations in each of three regions (input, grating, output) such that the tangential components of the solutions of neighbouring regions match at these interfaces. Phase information of the Fourier components of the refractive index profile given by Eq.\,\ref{eq:RefrIndexPatt} is inherently included in the RCWA. A set of coupled-wave equations is solved by calculating eigenvalues and eigenvectors of a matrix directly related to those Fourier components. The amplitudes at arbitrary diffraction orders are found by employing the boundary conditions.
In our work RCWA was used to fit the experimental DE defined as $\eta_s=I_s/I_{\text{\scriptsize{tot}}}$, with the diffracted intensity $I_s$ at the diffraction order $s$ and the sum of all diffracted intensities behind the sample $I_{\text{\scriptsize{tot}}}$, and yield the amplitudes and phases of the corresponding Fourier components at the index $s$ required to fully reconstruct $n(x)$.

% % % % % % % % 
\begin{figure}
\begin{center}
 \includegraphics[width=7.5cm]{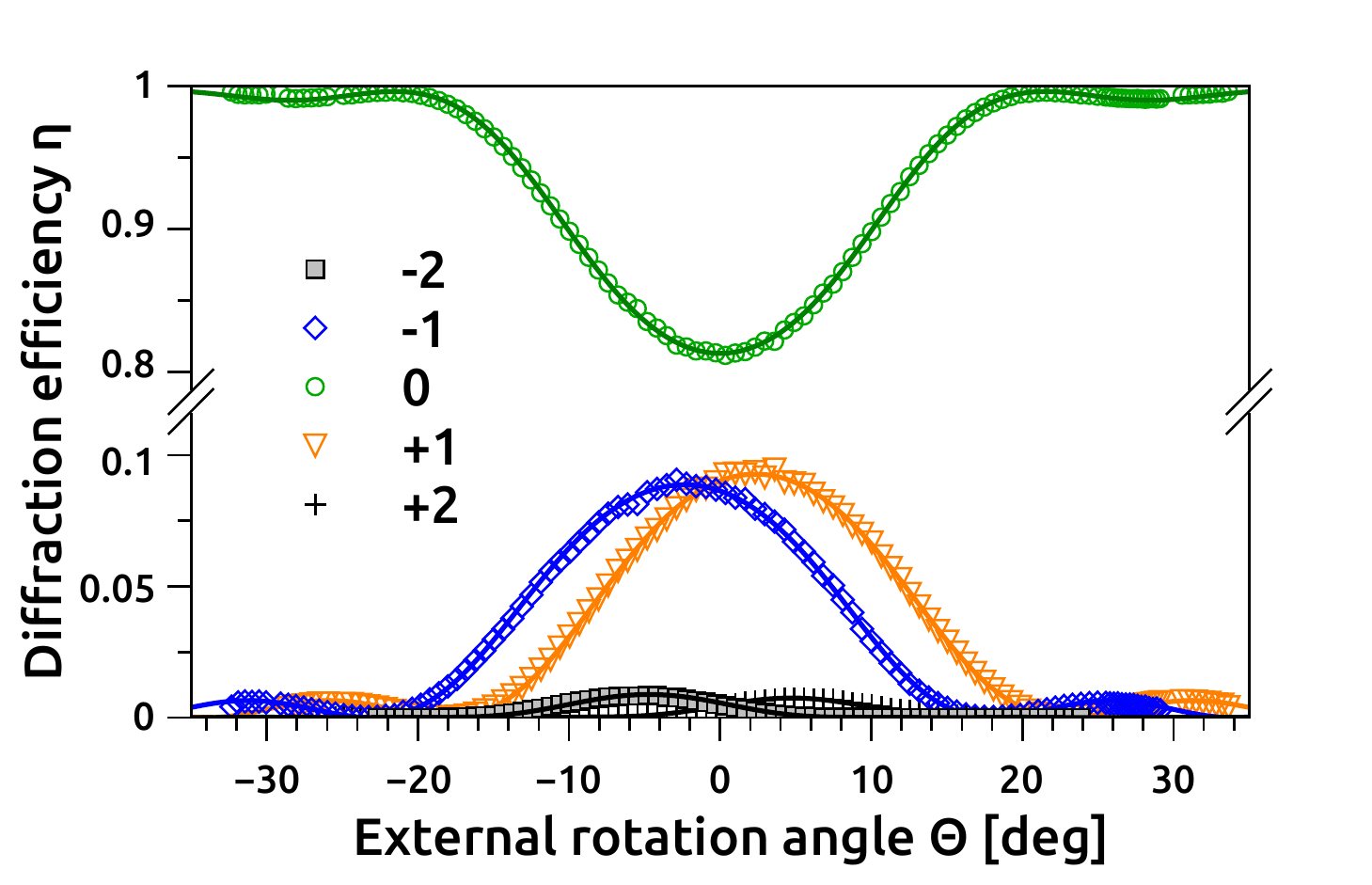}
\end{center}
\caption{\label{fig3} Measured angular dependence (data points) of the DE for the $\pm 2,\pm 1, 0$ orders for G1 at a wavelength of 633\,nm. Error bars are much smaller than the symbols. The $\pm 3$ orders were measured and included in the fitting procedure, but their DE is small and not shown here. Solid curves are RCWA fits to the data at seven diffraction orders, yielding $\varphi_2$ and $\varphi_3$ (see Table\,\ref{tab1}).} 
\end{figure}  

\begin{figure}
\begin{center}
 \includegraphics[width=7.5cm]{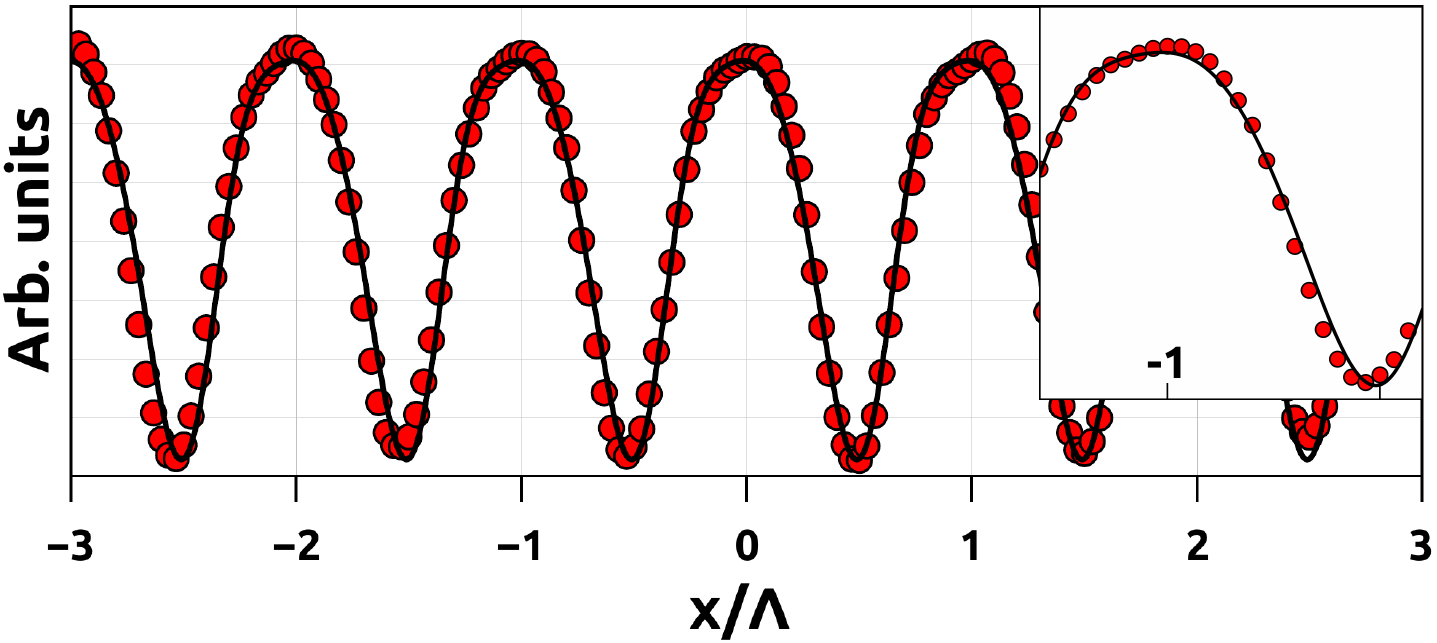}
\end{center}
\caption{
\label{fig4} 
Plotted are the data points as shown in Fig.\,\ref{fig1}\,(b) (filled, red circles). The refractive index profile of G1 from the RCWA fit-parameter estimation (as given in Table\,\ref{tab1}) was rescaled and shifted (solid, black curve) for comparison to the data. Peaks correspond to the polymer ridges of the structure. Inset: Third peak from the left, magnified. Even detailed features of the micrograph data are reproduced, as shown in the inset.
%Dashed line (blue): The same as before, but for $\varphi_2=0$ (c.f.\,dashed, blue lines in Fig.\,\ref{fig3}) instead of the experimentally obtained value $\varphi_2=\pi$ of Table\,\ref{tab1}. Note the narrow peaks and broad dips exhibited by the latter, in contrast to the data which is in accordance with the calculation for 
%$\varphi_2=\pi$.
} 
\end{figure}

To measure the angular dependence data, an s-polarized beam of a He-Ne laser ($\lambda=633$ nm) was used to observe diffraction signals from G1 being rotated about the $y$-axis (perpendicular to the plane of incidence) at angles $\Theta$ ($-55^\circ\ldots +55^\circ$), as is shown in Fig.\,\ref{fig2}\,(b).
Angular dependences of the diffracted intensities $I_s(\Theta)$ at $s=0,\pm 1,\pm 2,\pm 3$ were recorded by Si-photodiodes placed at the locations of the diffraction spots observed on a screen. The results are shown in Fig.\,\ref{fig3}.
Due to the small ratio $\lambda/\Lambda\approx 0.1$ in our experiment, the Bragg angles $\Theta_s=\pm\,3.6^\circ, \pm\,7.3^\circ, \pm\,10.9^\circ$ at $s=\pm 1,\pm 2,\pm 3$, respectively, are small, too. All seven observable diffraction orders overlapped within $\pm 20^\circ$ and exhibited sufficiently large DE to be readily detected. 
Solid curves in Fig.\,\ref{fig3} are least-squares fits to the RCWA. Fitting was performed including all data points. The RCWA is in excellent agreement with the experimental data, by which we can achieve accurate structure determination as is shown below: 
Estimations for the amplitudes $|n_1|,|n_2|$ and $|n_3|$ of the Fourier components as well as their phases $\varphi_2$ and $\varphi_3$ as obtained by the RCWA fit are given in the second column of Table\,\ref{tab1}. 
%Note that the $\varphi_s$ are \emph{relative} phases. 
The values of $\varphi_2$ and $\varphi_3$ were -- without loss of generality -- obtained with respect to $\varphi_1$ set to zero. The spatial refractive index profile as calculated from fit parameter estimations is shown in Fig.\,\ref{fig4} (solid, black curve). Since the refractive index modulation amplitude of a recorded NPC grating is proportional to $n_n - n_p$ \cite{suzukiAPL2002}, the \ce{SiO2} rich regions correspond to the low refractive index regions. Comparison with the data of Fig.\,\ref{fig1}\,(b) (filled, red circles in Fig.\,\ref{fig4}) makes the qualitative agreement obvious, thereby demonstrating the validity of our approach.
Even the slight asymmetry in the profile (slight flattening on the left side of each peak), which may be caused by spatially nonuniform lateral shrinkage during holographic exposure, is captured by the RCWA analysis of the diffraction data as can be seen from the inset in Fig.\,\ref{fig4}. 
 
\begin{table}
\begin{tabular}{|c|c|c|}\hline
grating & G1\,(light, $\lambda=633\,$nm) & G2\,(neutrons, $\lambda=1.7\,$nm) \\\hline
$|n_1|$ &$4.9382(37)\times 10^{-3}$ &  $2.592(28)\times 10^{-6}$ \\
$|n_2|$ &$1.072(15)\times 10^{-3}$ &  $5.03(58) \times 10^{-7}$ \\
$|n_3|$ &$1.72(48)\times 10^{-4}$
& \text{n.\,a.} \\\hline
$\varphi_1$ & $:=0$ & $:=0$ \\ 
$\varphi_2$ & $1.0581(39)\pi$ & $0.995(27)\pi$ \\ 
$\varphi_3$ & $0.37(15)\pi$ & \text{n.\,a.} \\ 
%$d$ ($\mu$m) & $13.34\pm 0.02$ & $114.2\pm0.8$ 
\hline

 \end{tabular}
\caption{\label{tab1} Amplitudes and relative phases for the Fourier coefficients of refractive index profiles for G1 and G2. The values were determined by RCWA fits to the data with $\varphi_1$ set to zero (see text). 
%The results for the thickness $d$ are in good agreement to the known thickness of spacers introduced between the carrier glass plates. 
Resulting profiles obtained from light and neutron diffraction measurements are shown in Figs.\,\ref{fig4} and \ref{fig6}, respectively.} 
\end{table}

Next, the refractive index profile of G2 ($\Lambda=1\,\mu\textrm{m}$, $d\approx 50\,\mu$m) is to be retrieved. Accurately resolving such structures by an optical microscope is possible but not straightforward, which underlines the potential impact of our solution to the phase problem, especially for refractive-index profiles, which have structural dimensions of the order of light wavelengths. 
In the case of G2, Kogelnik's theory \cite{kogelnikBSJ1969} allows to roughly estimate the expected peak width (the angular distance between the minima adjacent to the Bragg angle $\Theta_1$ at $s=1$ in an angular dependence plot of DE) as $2 n_0 \Lambda/d\approx 3.4^\circ$ while one obtains $\Theta_1\approx 17^\circ$ at $\lambda=633$\,nm and $\Theta_1\approx 10^\circ$ at a UV wavelength of $\lambda=351\,$nm, say. Thus, it is difficult to produce peak overlap to transfer the diffraction process from the Bragg regime to the multi-wave interference regime for the determination of the refractive index profile of G2, by use of table-top laser light sources. However, a key point of our approach is that depending on the particular material class investigated, there might be other kinds of radiation available for diffraction experiments to obtain data similar to the ones shown in Fig.\,\ref{fig3}. For instance, since the refractive index profile is produced by the density modulation of one of the involved material components (nanoparticles, in our case), it is known that small angle neutron scattering (SANS; see, for instance, \cite{willisBook2009}) provide well-established tools \cite{fallyPRL2010}: For a typical de Broglie wavelength of neutrons in a SANS experiment of 1\,nm, say, one may expect Bragg angles $\Theta_1\approx \lambda/(2\Lambda)$ of the order of $0.03^\circ$, which can be detected with state-of-the-art SANS instruments, thanks to long sample-detector distances (up to 20\,m) and sufficient spatial resolution of detectors. Thus, considerable peak-overlap for G2 and, therefore, multi-wave interference is achievable with neutrons.

\begin{figure}
\begin{center}
 \includegraphics[width=7.8cm]{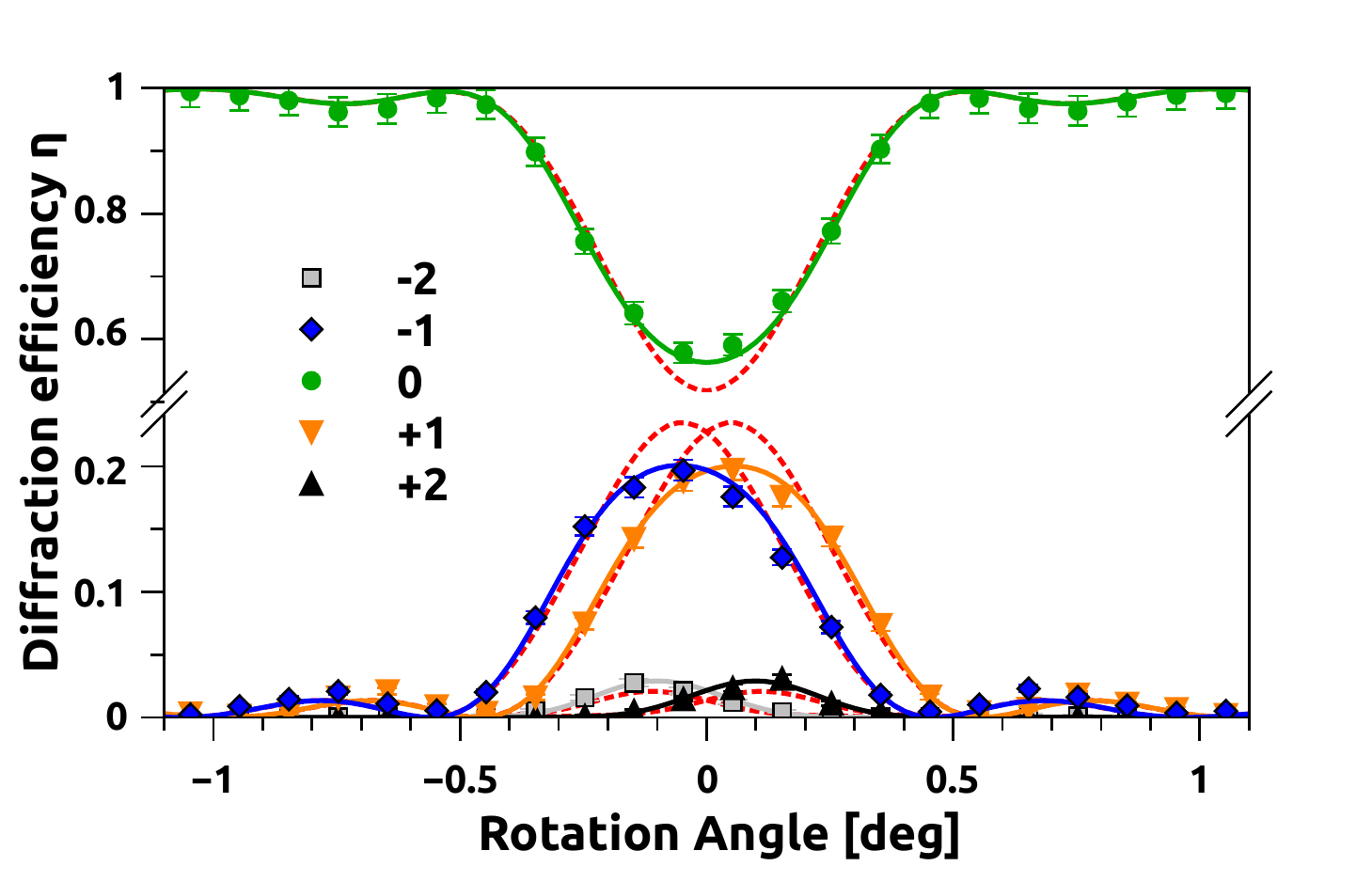}
\end{center}
\caption{\label{fig5} Measured angular dependence (data points) of DE for G2 at the $\pm 2,\pm 1, 0$ orders and at the neutron wavelength of $1.7$\,nm. Five-wave interference was observed near the normal incidence. An RCWA fit (solid curves) again allows to retrieve amplitudes and phases of the grating's Fourier components. The resulting fit parameter estimation is given in Table\,\ref{tab1}. To illustrate the importance of the correct phase value, simulations for 
$\varphi_2=0$ (dashed, red curves) instead of the experimentally obtained value $\varphi_2=0.995\,\pi$ are added.}
\end{figure}

\begin{figure}
\begin{center}
 \includegraphics[width=7.9cm]{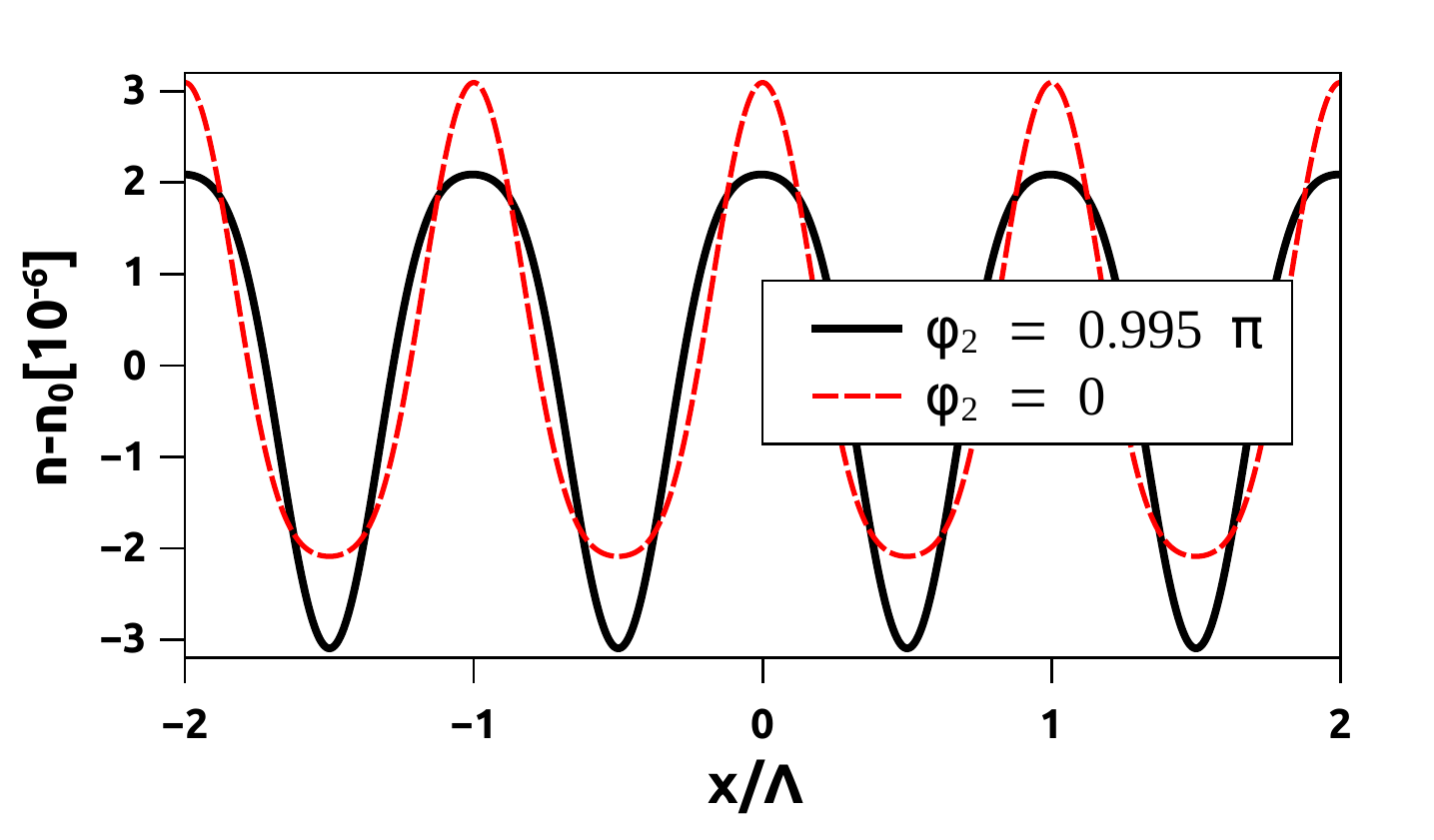}
\end{center}
\caption{\label{fig6} The neutron-refractive index profile of G2 calculated from the RCWA fit-parameter estimation as given in Table\,\ref{tab1} (solid, black curve). 
The neutron-refractive index profile of G2 is also plotted for 
$\varphi_2=0$ (dashed, red curve) instead of the experimentally obtained value $\varphi_2=0.995\,\pi$. }
\end{figure}

The neutron experiment was performed by use of the instrument SANS-I of the SINQ neutron source of Paul-Scherrer Institute in Villigen, Switzerland. Cold neutrons at a mean wavelength of $1.7$\,nm (width of wavelength distribution $\Delta\lambda/\lambda\approx10\%$) were diffracted from G2. The beam divergence was limited to about 1\,mrad, using collimation slits. The diffracted intensities were measured using a two-dimensional detector of $7.5\times7.5\,$mm$^2$ pixel size. To adjust the peak width (estimated by $2\Lambda/d$, see our above discussion) and the peak height for our purpose, G2 was tilted around its grating vector by $\zeta\approx60^\circ$ which increases the effective thickness to $d\rightarrow d/\cos\zeta$, i.\,e. by a factor of two \cite{somenkovSSC1978,fallyPRL2010}. The results are shown in Fig.\,\ref{fig5}. Similarly to the DE of G1 shown in Fig.\,\ref{fig3}, the observed diffraction took place in the multiple-wave interference regime, that is, many (five, in this case) diffraction orders are observable within the angular range of $\Theta=\pm 0.25^\circ$. Applying the RCWA fit, we were, again, able to extract the full information (amplitudes \emph{and} phases) of the grating's Fourier components up to the 2nd-order. They are given 
in Table\,\ref{tab1} (third column). Simulations of the DE assuming $\varphi_2=0$ instead of the experimentally obtained value are also plotted (red, dashed curves) in Fig.\,\ref{fig5}. The disagreement of simulation at $\varphi_2=0$ and data is easily resolved in a standard SANS experiment. The neutron-refractive index profile of G2 -- corresponding to the spatial density distribution of formed polymer and nanoparticles -- as calculated from the RCWA parameter estimation is shown in Fig.\,\ref{fig6} together with its counterpart, assuming $\varphi_2=0$, for comparison. 
Note that the refractive index modulation amplitude for neutrons is proportional to 
$-(b_n\rho_n - b_p\rho_p)$, 
where $b_n$ ($b_p$) and 
$\rho_n$ ($\rho_p$) are the mean bound coherent scattering length \cite{willisBook2009} and the atomic density of nanoparticles (the formed polymer), respectively. 
Since $b_n\rho_n$ is larger than $b_p\rho_p$ for \ce{SiO2} nanoparticles \cite{fallyPRL2010}, the \ce{SiO2} rich regions correspond to lower refractive index regions in Fig.\,\ref{fig6}, as similar to Fig.\,\ref{fig4}. No third order component was measurable for G2 and therefore the profile in Fig.\,\ref{fig6} is symmetric, in contrast to the profile of G1. We attribute this difference to the mutual diffusion process, which is much more significant in G2 due to the smaller grating spacing.

Of course, it is -- in many cases -- more convenient to apply optical microscopy for the estimation of the refractive-index profile, but resolution limits of this method are laborious to overcome in the range of typical structure constants of several hundred nanometers. Furthermore, electron microscopy and physico-chemical analyses often depend on sample preparation techniques -- producing thin slices or breaking samples to look at surfaces -- that are cumbersome or even unreliable in the sense that they could, in the worst case, mechanically alter the structure to be investigated. Our technique, however, can provide sufficient resolution in bulk so long as a suitable state-of-the-art instrument, meeting the wavelength requirements to work in multi-wave interference regime for a structure of interest, can be accessed at one of the many facilities worldwide. Resolution is limited by the --usually excellent -- detector sensitivity and background suppression necessary for reliable observation of weak higher-order diffraction signals. The RCWA analysis (programmed in the Python programming language) carried out here on an off-the-shelf PC can be applied to more complicated structures by using more powerful computation infrastructure, nowadays available at many institutions.

Finally, we note that the technique presented here is closely related to previous proposals to employ multi-wave interference (see, for instance, \cite{chapmanPRL1981,changPRL1982,woolfsonBook1995,shenPRL1998}). The fundamental difference between those and our approach is, however, that apart from a careful choice of wavelengths, only standard diffraction procedures are applied here. No extra data needs to be taken. Our approach can be seen as generalization of what is discussed in the recent Ref.\,\cite{heacockIUCrJ2020}.

In summary, we have demonstrated the determination of refractive-index profiles by diffraction from one-dimensional holographic phase gratings recorded in \ce{SiO2} nanoparticle-dispersed NPC films. We have shown that full phase retrieval can be made without the need for extra data collection schemes. In particular, by choosing the probe wavelength some orders of magnitude shorter than the structural dimensions being investigated, Bragg regime diffraction can be turned into diffraction in the regime of multi-wave interference for a given sample. Fitting of a multi-wave coupling model such as the well-known RCWA to the resulting angular dependence data allows for the accurate determination not only of amplitudes but also of phases of the Fourier components of generic profiles of refractive index modulation. When dealing with photonic structures (like gratings) it is a common practice to avoid multi-wave interference for ease of the analysis, in particular when it comes to the evaluation of the underlying parameters. Here, we have shown that getting rid of this habit and embracing the usually shunned multi-wave interference regime can pay off at the prize of a relatively low increase in experimental and analytical complication. Our technique is a purely physical solution to the phase problem of diffraction. 

We would like to thank G. Heuberger for preliminary experimental work. J.\,G. and Y.\,T. would like to acknowledge the financial supports by Ministry of Education, Culture, Sports, Science, and Technology of Japan under grant 
No.\,25-03052 and No.\,15H03576.

%\bibliography{holo,vo,neutron,publications_url,prbstr,/home/juergen/Documents/laTex/localtexmf/bibtex/bib/juergen}
\end{document}